\documentclass[default]{sn-jnl}

\usepackage{graphicx}%
\graphicspath{{figures/}}
\usepackage[squaren]{SIunits}
\usepackage{multirow}%
\usepackage{amsmath,amssymb,amsfonts}%
\usepackage{amsthm}%
\usepackage{mathrsfs}%
\usepackage[title]{appendix}%
\usepackage{xcolor}%
\usepackage{textcomp}%
\usepackage{manyfoot}%
\usepackage{booktabs}%
\usepackage{algorithm}%
\usepackage{algorithmicx}%
\usepackage{algpseudocode}%
\usepackage{listings}%
\usepackage{hyphenat}
\usepackage[capitalise]{cleveref}

\theoremstyle{thmstyleone}%
\theoremstyle{thmstyletwo}%

\theoremstyle{thmstylethree}%

\definecolor{red}{rgb}{1,0,0}

\begin{document}

\title{Practical security of twin-field quantum key distribution with optical phase-locked loop under wavelength-switching attack}


\author[1]{\fnm{Qingquan} \sur{Peng}}

\author[2]{\fnm{Jiu-Peng} \sur{Chen}}

\author[1]{\fnm{Tianyi} \sur{Xing}}

\author[1]{\fnm{Dongyang} \sur{Wang}}

\author[1]{\fnm{Yizhi} \sur{Wang}}

\author*[2]{\fnm{Yang} \sur{Liu}}\email{liuyang@jiqt.org}

\author*[1]{\fnm{Anqi} \sur{Huang}}\email{angelhuang.hn@gmail.com}

\affil[1]{\orgdiv{Institute for Quantum Information \& State Key Laboratory of High Performance Computing, College of Computer Science and Technology}, \orgname{National University of Defense Technology}, \orgaddress{\city{Changsha}, \postcode{410073}, \state{Hunan}, \country{China}}}

\affil[2]{\orgname{Jinan Institute of Quantum Technology and Hefei National Laboratory Jinan Branch}, \orgaddress{\city{Jinan}, \postcode{250101}, \state{Shandong}, \country{China}}}


\abstract{The twin-field class quantum key distribution (TF-class QKD) has experimentally demonstrated the ability to surpass the fundamental rate-distance limit without requiring a quantum repeater, as a revolutional milestone. In TF-class QKD implementation, an optical phase-locked loop (OPLL) structure is commonly employed to generate a reference light with correlated phase, ensuring coherence of optical fields between Alice and Bob. In this configuration, the reference light, typically located in the untrusted station Charlie, solely provides wavelength reference for OPLL and does not participate in quantum-state encoding. However, the reference light may open a door for Eve to enter the source stations that are supposed to be well protected. 
Here, by identifying vulnerabilities of an acousto-optic modulator (AOM) in the OPLL scheme, we propose and demonstrate a wavelength-switching attack on a TF-class QKD system. 
This attack involves Eve deliberately manipulating the wavelength of the reference light to increase mean photon number of prepared quantum states, while maintaining stable interference between Alice and Bob as required by TF-class QKD protocols. The maximum observed increase in mean photon number is 8.7\%, which has been theoretically proven to compromise the security of a TF-class QKD system. Moreover, we have shown that with well calibration of the modulators, the attack can be eliminated. Through this study, we highlight the importance of system calibration in the practical security in TF-class QKD implementation.}

\maketitle

\section{Introduction}\label{sec:intro}

Quantum key distribution (QKD), based on the principles of quantum physics, enables two remote parties to securely share a secret key over a public and insecure channel, demonstrating information-theoretic security irrespective of an eavesdropper's computational power~\cite{bennett2014}. Over the past approximately four decades, QKD technology has progressively evolved and matured to such an extent that its practical deployment in optical-fiber networks is now feasible~\cite{xu2020}. Throughout this advancement, the primary objectives have been to achieve extended transmission distances and higher key rates while ensuring robust security performance. However, accomplishing each of these aforementioned objectives presents significant challenges.

Channel loss is a crucial barrier of enhancing the key rate and transmission distance of QKD systems. This is primarily due to the vulnerability of single photons, which are used for encoding quantum keys, to scattering and absorption within the transmission channel. However, amplification of these photons is not feasible~\cite{archana2015}, resulting in reduced detection probabilities over extended distances. In QKD systems without incorporating quantum repeaters, there exists a fundamental limit on the secure key rate known as the Pirandola-Laurenza-Ottaviani-Bianchi (PLOB) bound~\cite{pirandola2017}. This bound scales linearly with channel transmittance $\eta$, denoted as $O(\eta)$~\cite{takeoka2014}.

Fortunately, a recent protocol called twin-field QKD (TF QKD) has been proposed, surpassing the fundamental rate-distance limit and revolutionizing the scalability of both key rate and channel loss to be $O(\sqrt{\eta})$~\cite{lucamarini2018}. TF QKD leverages the advantage that two distant peers, Alice and Bob, transmit two optical fields with correlated phase to conduct single-photon interference at an intermediate station named Charlie. This implies that only one photon from either Alice or Bob needs to arrive and be detected at Charlie in order to generate a secret key, while the other peer transmits vacuum states that are inherently immune to channel loss. In addition to overcoming transmission limits, TF QKD also inherits the advantages of measurement-device-independent (MDI) QKD~\cite{lo2012,curty2014}, where Charlie can be fully untrusted as an intermediate station. The security of TF QKD against general attacks has been theoretically proven, establishing a solid foundation for its implementation~\cite{ma2018,wang2018,cui2019,curty2019,yin2019,currsLorenzo2021}. Following this theoretical development, experimental demonstrations have shown that TF-class QKD outperforms the PLOB bound~\cite{minder2019,liu2019a,chen2020,zhong2021,pittaluga2021,liu2021,chen2022,wang2022,liu2023}, achieving record-breaking distances up to $1002~\kilo\meter$~\cite{liu2023,liu2023a}. These achievements highlight the feasibility of implementing intercity QKD without relying on trusted relays.

However, the practical security performance, another crucial aspect of QKD, has not yet been investigated in the context of this novel TF-class QKD. Regarding its implementation, it is quite challenging to generate identical optical fields with correlated phase by two separate parties as required by the protocol. This unique requirement introduces a distinct implementation scheme that may also introduce unforeseen security vulnerabilities. In initial proof-of-principle experiments~\cite{lucamarini2018,zhong2019}, the light source was placed at an untrusted location and transmitted to Alice and Bob. Unfortunately, this approach is evidently insecure since Eve can manipulate the quantum signal encoding key information by controlling the light source. To mitigate this vulnerability and strictly adhere to the structure of TF QKD protocol, subsequent experiments equipped each Alice and Bob with separate lasers. However, this mitigation strategy gives rise to a new challenge - ensuring that Alice's and Bob's lasers produce optical fields with identical phase correlation. To address this concern, achieving wavelength consistency of the two separate laser sources is accomplished through the laser injection technique. In this technique, Charlie applies a seed laser to independently inject into Alice's and Bob's laser sources~\cite{fang2020, liu2021}. However, this scheme potentially introduces a vulnerability for Eve to gain access to the assumed-to-be-protected source station~\cite{shor2000, renner2008} - Eve could manipulate the seed light via Charlie or exploit the public channel for similar laser-seeding attacks~\cite{sun2015,huang2019,pang2020}. 
Moreover, TF QKD schemes might also be susceptible to Trojan-horse attacks and induced-photorefractive attacks~\cite{lu2023a,ye2023}.
Therefore, ensuring practical security in TF-class QKD systems is more intricate compared to previous QKD implementations due to their unique realization schemes and techniques that require heightened attention.

In addition to the aforementioned known threats on these schemes, an advanced technique known as optical phase-locked loop (OPLL) has been extensively employed in recent implementations to achieve persistent stable interference by generating identical optical fields between Alice and Bob \cite{minder2019, wang2019, liu2019a,chen2020,pittaluga2021, liu2021,chen2022,wang2022,liu2023,chen2024}. In this approach, Alice's and Bob's lasers are synchronized with a reference laser source for both frequency and phase \cite{liu2021,chen2022,wang2022,liu2023}. While this method enhances the stability of correlated phases in two optical fields and supports the realization of a secure interference TF-class QKD implementation, its practical security aspects remain unexplored.

In this study, our focus lies on the practical security performance of a TF QKD system employing the OPLL scheme. In this scheme, although the reference light is not directly seeded into Alice's/Bob's laser, each source station does provide an entry to enable transmission of the reference light into the theoretically protected space, interfering with Alice's/Bob's laser beam. The OPLL then acts as feedback to adjust the wavelength of Alice's/Bob's laser for mutual matching, ensuring identical optical fields with correlated phase. Our investigation reveals a loophole in wavelength-intensity correlation in the OPLL scheme. By exploiting this loophole, Eve can switch wavelengths of the reference light, thereby controlling its wavelength and indirectly affecting the mean photon number of quantum states prepared by Alice and Bob, while maintaining stable interference between them. This attack is referred to as a wavelength-switching attack. Surprisingly, under this scenario of a wavelength-switching attack, there is an increase in mean photon number by 8.7\%, leading to insecurity of secret key.


\section{Results}\label{sec:results}

\subsection{The idea of wavelength-switching attack}
\label{sec:hacking}

The fundamental mechanisms of TF-class QKD are elucidated in this section, providing insights into its superiority over the PLOB bound. Subsequently, the OPLL, as the pivotal component of TF-class QKD systems, is introduced in detail. Building upon this foundation, a specialized hacking method targeting TF-class QKD systems equipped with OPLL is proposed as the wavelength-switching attack.

\begin{figure*}[htbp]
	\centering
	\includegraphics[width=1\textwidth]{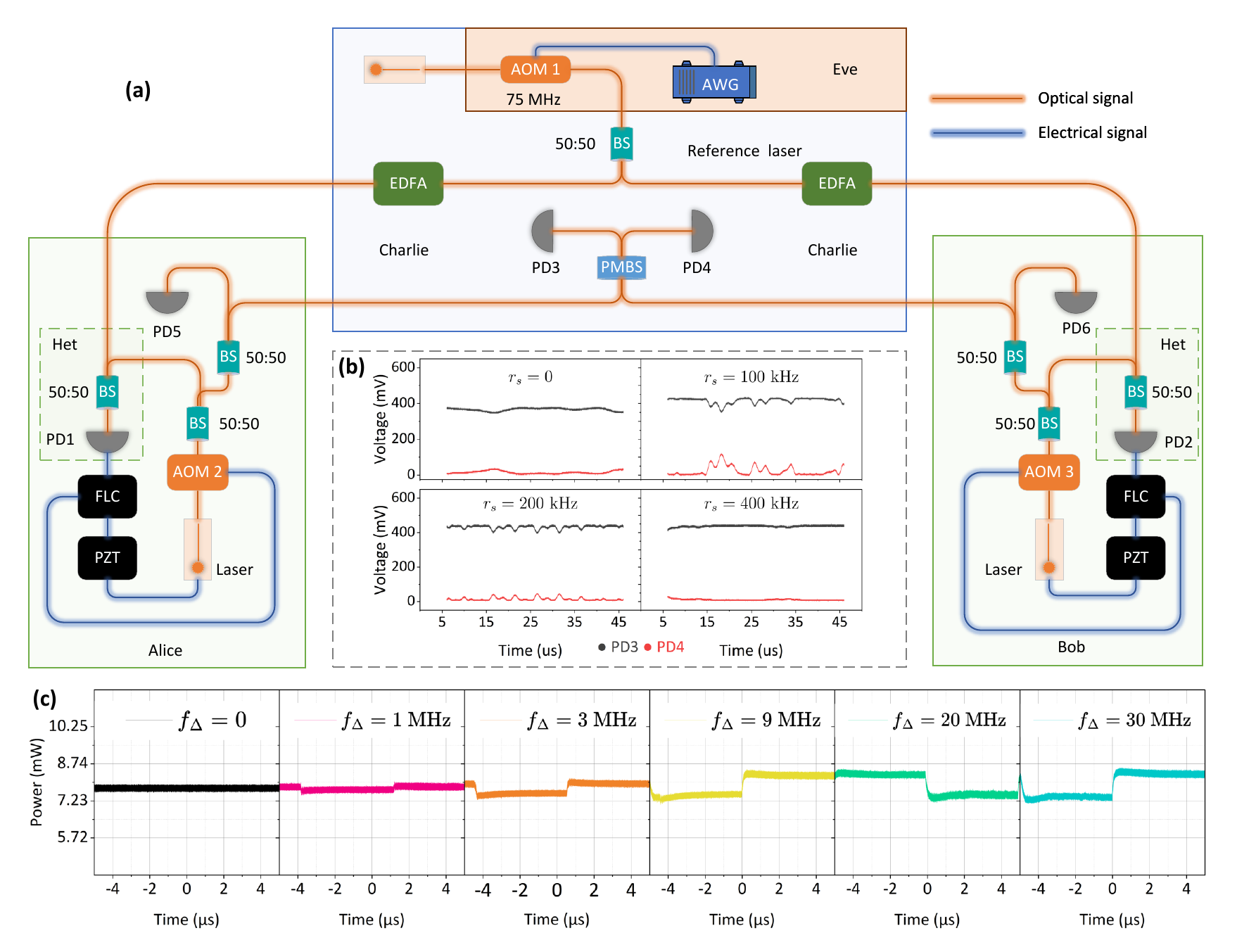}
	\caption{Experimental schematic diagram and testing results.
	(a) Schematic diagram illustrating the structure of the OPLL in a TF-class QKD system, along with the corresponding wavelength-switching attack. The red and blue connecting lines represent optical fiber and electrical cables, respectively. The frequency-locking controller~(FLC) modulates the operation of Alice's/Bob's piezoelectric transducer (PZT) and acousto-optic modulator (AOM2/AOM3) to ensure the coherence of optical field between Alice and Bob. The optical intensity of Alice's/Bob's laser is measured by photodetector PD5/PD6, whose bandwidth is $500~\mega\hertz$, connected to an oscilloscope with sampling rate of 80 GSa/s. EDFAs, erbium-doped fiber amplifiers; BS, beam splitter; Het, heterodyne measurement; PMBS, polarization-maintaining beam splitter.
	(b) Interference measurements obtained by Charlie in the TF QKD system under both normal operation conditions and during the presence of a wavelength-switching attack at $f_\Delta=30~\mega\hertz$.
	(c)  Optical power measurements recorded at AOM2's output while subjected to a wavelength-switching attack with the rate of wavelength switching $r_s=100~\kilo\hertz$.
	}
	\label{fig:tfqkdscheme}
\end{figure*}

\subsubsection{TF-class QKD based on OPLL}
\label{subsec:tfqkd}

The TF-class QKD protocol is an efficient variant of the MDI QKD protocol. In the original MDI QKD protocol, Alice and Bob each transmit a photon to Charlie for a two-photon Bell state measurement. In contrast, the TF-class QKD protocol achieves post-selected entanglement by performing single-photon interference at Charlie's side instead of a two-photon Bell state measurement, enabling it to achieve a key rate proportional to the square root of channel transmittance ($O(\sqrt{\eta})$). To ensure successful single-photon interference at Charlie's side, maintaining high coherence in the optical fields emitted by Alice and Bob becomes crucial. Consequently, ensuring wavelength consistency between Alice's and Bob's light sources becomes essential in implementing the TF-class QKD protocol.

To achieve correlated phase between two optical fields, Alice and Bob employ the OPLL structure in TF-class QKD systems for wavelength and phase synchronization, as depicted in~\cref{fig:tfqkdscheme}(a). The OPLL comprises a frequency-locking controller (FLC), piezoelectric transducer (PZT), acousto-optic modulator (AOM), and heterodyne measurement. Charlie utilizes an ultra-stable laser to generate a reference light, which is subsequently transmitted to Alice and Bob. This reference light does not carry any encoding information but can be amplified through erbium-doped fiber amplifiers (EDFAs) during transmission. Upon receiving the reference light, Alice/Bob employs heterodyne measurement to determine the frequency/phase difference between her/his laser and the reference light. Specifically, a photodetector and phase discriminator detect the frequency/phase deviation, which is then processed by a proportional integrator (PI) controller to adjust Alice's/Bob's laser frequency/phase to match that of the reference light. This adjustment is achieved by controlling PZTs and two inserted AOMs in their respective lasers. The PZTs are used for adjusting initial frequencies and constraining slow frequency drifts while AOMs compensate for fast frequency shifts. When there is a change in the wavelength of the reference light, the FLC rapidly adjusts the frequency modulation via the AOM. Following this, to maintain the AOM in their optimal operating state, the FLC gradually alters the frequency modulation of the AOM back to its central value, while concurrently modifying the PZT to compensate for the frequency discrepancy. The coordinated operation of PZTs and AOMs enables precise frequency synchronization between Alice's and Bob's lasers in TF-class QKD systems. 

\subsubsection{The working principle of wavelength-switching attack}

As previously mentioned, in TF-class QKD systems, the OPLL structure has been employed to ensure the coherence of optical fields between Alice and Bob \cite{liu2019a,wang2019, chen2020,liu2021,chen2022,wang2022,liu2023}. However, the implementation of this structure introduces a potential vulnerability by creating an open port for the reference light to enter Alice's/Bob's lab under the assumption that it is well protected. This reference light could serve as a pathway for Eve to manipulate the internal devices of Alice and Bob, thereby compromising one of the fundamental assumptions of QKD - strict protection of internal devices. In line with addressing practical security concerns, this subsection presents how Eve can exploit this open port for manipulating Alice's and Bob's laser sources through a wavelength-switching attack aimed at compromising the security of TF-class QKD.

In the TF-class QKD system, exemplified by~\cref*{subsec:tfqkd}, the OPLL structure facilitates precise adjustment of Alice's or Bob's laser wavelength through coordinated actions of AOM and PZT. Our primary focus lies in elucidating the operational procedure of OPLL and exploiting a loophole in the AOM within this configuration to control the output intensities of quantum states prepared by Alice and Bob. Specifically, we observe that any sudden fluctuations in wavelength trigger prompt corrective shifts by the AOM. However, an unforeseen imperfection arises from the dependence of insertion loss on modulation frequency $f_{aom}$, which is further illustrated in~\cref{App:AOM}.

Based on the aforementioned concept, we propose a wavelength-switching attack for the TF-class QKD system, as depicted by the red area in~\cref*{fig:tfqkdscheme}(a).
In this attack scenario, a wavelength-switching module consisting of an AOM1 and an arbitrary waveform generator (AWG) is inserted at the output of the reference laser, as shown in the red box in Fig.~\ref{fig:tfqkdscheme}(a).
Consequently, the AOM induces frequency shifts between $f_{aom1}$ and $f_{aom2}$, while the AWG controls the rate of wavelength switching ($r_s$). By carefully adjusting these parameters, Eve ensures that only AOM operates without any adjustment on PZT in Alice's/Bob's OPLL module. Subsequently, she transmits a reference light with a pattern of switched wavelengths to both Alice and Bob. Upon receiving this reference light, Alice and Bob measure its heterodyne frequency relative to their local light source. The measurement reveals two peak values corresponding to different wavelengths of the reference light, prompting FLC to adjust AOM levels in Alice/Bob accordingly. In essence, Eve manipulates the wavelength of reference light to trigger wavelength switching in Alice's/Bob's local light sources. Importantly, this wavelength switching also reduces insertion loss in AOMs leading to increased intensity of quantum states prepared by Alice/Bob. If not monitored properly by Alice and Bob through their output light signals, this increased intensity could potentially enable Eve to eavesdrop on their secret key.

It should be noted that in a wavelength-switching attack, variations in insertion loss of AOM are determined by $f_{\Delta}= |f_{aom1} - f_{aom2}|$. Hence, when selecting a value for $f_\Delta$, Eve must ensure it is sufficiently large to cause a discernible change in mean photon numbers within quantum states generated by Alice and Bob. Moreover, adjustments to $f_\Delta$ are constrained within limits defined by FLC and AOM's maximum capabilities within OPLL. Subsequently, an experimental demonstration illustrating this wavelength-switching attack will be presented in order to emphasize its potential threat arising from unauthorized light infiltration into TF-class QKD source stations.

\subsection{Experimental demonstration}
\label{sec:experiment}

In this wavelength-switching attack, Eve exploits the security vulnerability of the AOM in the OPLL module. Therefore, this section initially presents the calibrated characteristics of the AOM. Based on these calibration results, we determine the parameters required to conduct a wavelength-switching attack and demonstrate its impact on TF-class QKD later in this section.

\subsubsection{The characteristics of the AOM}
\label{subsec:aom}

\begin{figure}[h]
	\centering
	\includegraphics[width=0.5\textwidth]{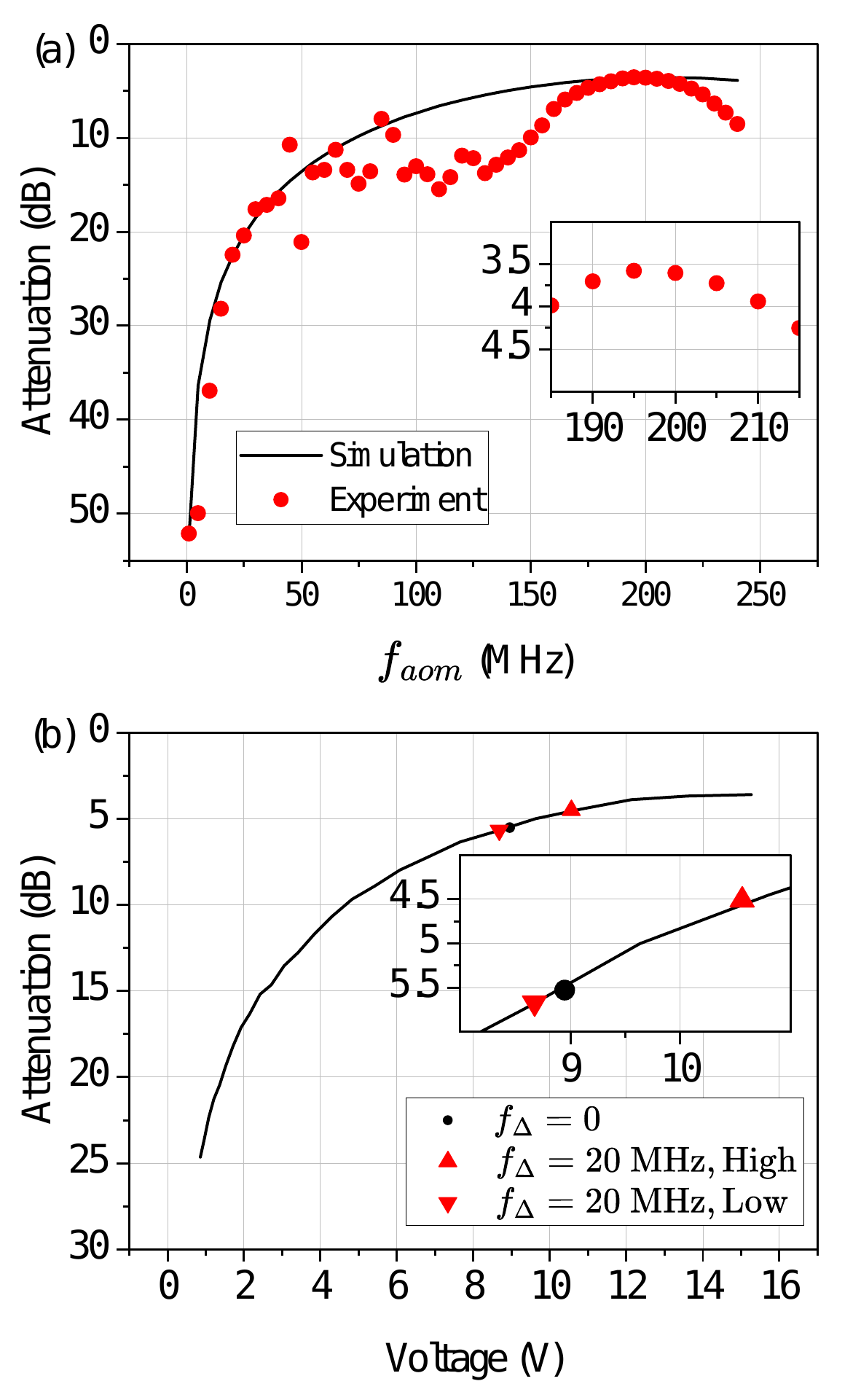}
	\caption{
		Characterization of the AOM. 
(a) The relationship between AOM attenuation and frequency modulation was characterized.
(b) The relationship between AOM attenuation and driving voltage was characterized. The red upward/downward triangles represents that the driving voltage of AOM is $10.56~\volt$/ $8.68~\volt$.
The input power of the AOM was $6.22~\milli\watt$, and its model number is GST200.
	}
	\label{fig:attenuation}
\end{figure}

To investigate the characteristics of the AOM, a comprehensive test was conducted to examine its response under various input conditions, with a specific focus on two key dimensions: carrier frequency and driving voltage. Regarding the test on carrier frequency, a constant driving voltage of $15.84~\volt$ was applied to the AOM, and a sinusoidal driving signal with frequencies faom ranging from $0$ to $240~\mega\hertz$ was injected into it. The initial laser wavelength, $\lambda_{\rm{in}}=v/f_{\rm{in}}=1550~\nano\meter$ where $v$ represents the speed of light in optical fiber and $f_{\rm{in}}$ is the frequency. Upon receiving the driving signal, the AOM effectively modulates the light's wavelength to $\lambda_{out} = v/\left( f_{\rm{in}} + f_{\rm{aom}} \right)$. By continuously varying $f_{aom}$ during the experiment, accurate characterization of how different values of $f_{aom}$ corresponded to changes in output power was obtained. During testing on driving voltage, a fixed-frequency sinusoidal driving signal at $200$~$\mega\hertz$ was applied to the AOM while systematically adjusting peak-to-peak voltages within a range of  $1$~$\volt$ to $15$~$\volt$. This meticulous adjustment allowed for establishing an empirical relationship between variations in input voltage levels and corresponding changes in output efficiency for precise performance evaluation.

The relationship between AOM attenuation and frequency modulation is illustrated in~\cref{fig:attenuation}(a). In this subfigure, the black solid curve represents simulation results, while the red dots correspond to experimental measurements. The observed attenuation spans a range from over $50~\deci\bel$ to a few decibels as a function of modulation frequency. Notably, the minimum measured attenuation value for the AOM employed in this study is $3.57~\deci\bel$, occurring at a modulation frequency of $195~\mega\hertz$. This variation in attenuation primarily stems from the effective coupling of light within the AOM device and its piezoelectric transducer efficiency. Additionally, imperfect coupling may contribute to discrepancies between experimental data (red dots) and simulation results (black curve). Disregarding coupling efficiency reveals that higher modulation frequencies result in increased ultrasonic power, directly influencing output efficiency. For comprehensive details on the working principle and model of AOM, please refer to ~\Cref*{App:AOM}.

The relationship between AOM attenuation and driving voltage is depicted in~\cref{fig:attenuation}(b). In the subfigure, the solid black curve represents the experimental measurements. The attenuation varies from over $25~\deci\bel$ to a few decibels with fluctuations in the driving voltage. The minimum observed attenuation value for the utilized AOM in this study is $3.59~\deci\bel$, corresponding to a driving voltage of $15~\volt$, and modulation frequency of $200~\mega\hertz$. The fluctuation in attenuation primarily arises due to insufficient excitation of the piezoelectric transducer caused by decreasing voltage, resulting in inadequate ultrasonic power output.

Thus, in the TF-class QKD system under wavelength-switching attack, the attenuation of an AOM is influenced by both frequency modulation and drive voltage. For instance, when $f_{aom1} = 200~\mega\hertz$, $f_{aom2} = 180~\mega\hertz$ and, thus, $f_{\Delta}=20~\mega\hertz$, \cref{fig:attenuation} (a) shows that frequency modulation results in an increase in attenuation by $0.69~\deci\bel$. Meanwhile, \cref{fig:attenuation}(b) illustrates that the driving voltage is increased from $8.95~\volt$ (black dot) to $10.56~\volt$ (red upward triangle, high voltage), resulting in a $1.02~\deci\bel$ decrease in attenuation. The combined effect of these two factors ultimately leads to an overall attenuation reduction of $0.33~\deci\bel$ for the AOM, which is consistent with the measured value, as indicated in Fig.~2(c) when $f_\Delta = 20~\mega\hertz$. When the AOM2 (AOM3) work frequency returns to $200~\mega\hertz$, \cref{fig:attenuation}(b) also illustrates that the driving voltage reduces to $8.68~\volt$  (red downward triangle, low voltage), resulting in a $0.14~\deci\bel$ increase in attenuation. Demonstration of wavelength-switching attack with more details are presented in the following subsections. 

\subsubsection{Testing of wavelength-switching attack}

\begin{figure}[h]
	\centering
	\includegraphics[width=0.5\textwidth]{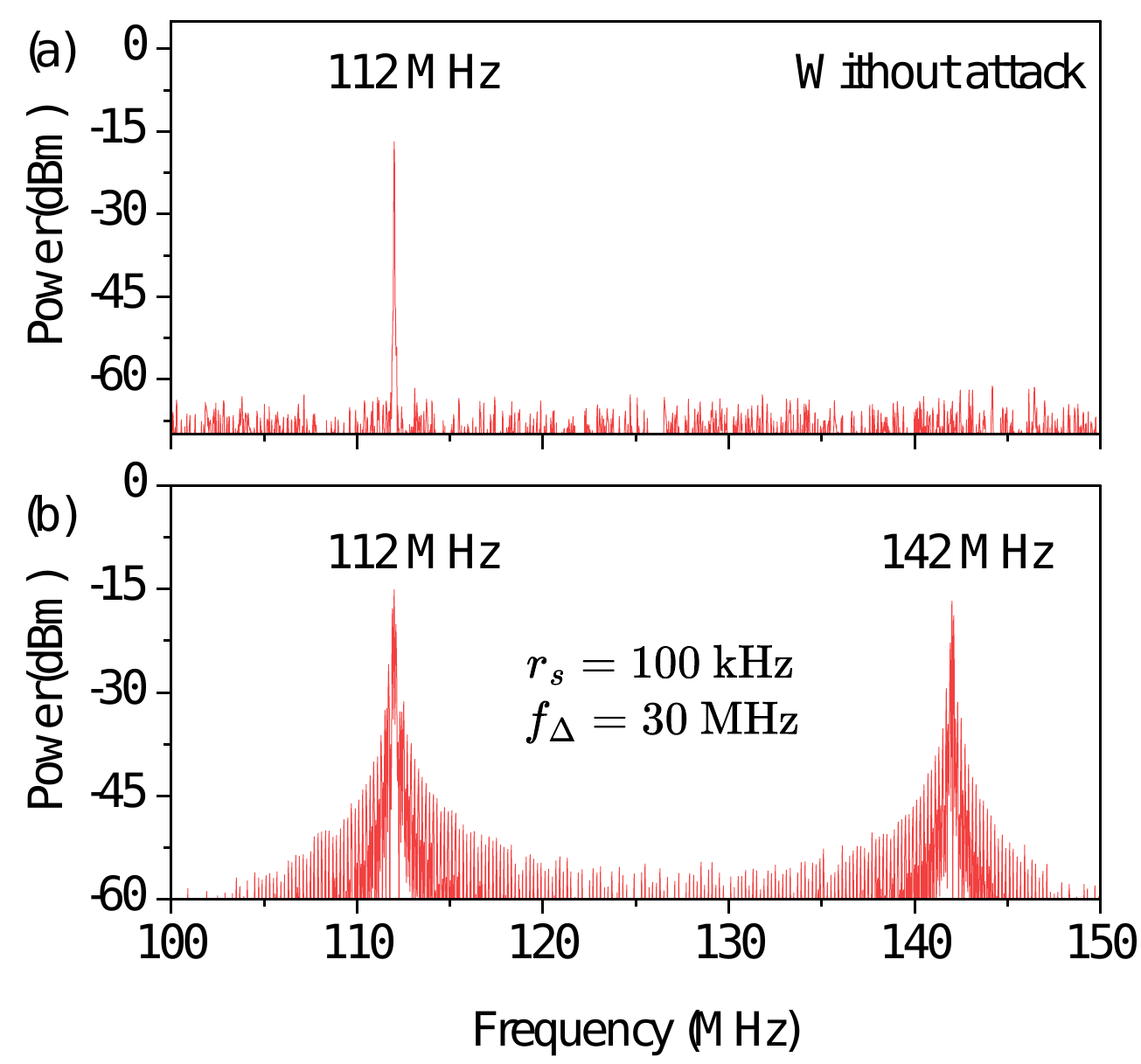}
	\caption{
		The heterodyne frequency experimental results obtained from PD1 in the testing scheme depicted in~\cref{fig:tfqkdscheme}(a) are presented. (a) The measured frequency of heterodyne distribution without any attack is reported. (b) The measured frequency of heterodyne distribution under attack conditions, with $f_{\Delta} = 30~\mega\hertz$ and $r_s=100~\kilo\hertz$, is also provided.
	}
	\label{fig:detectpd1}
\end{figure}

In the tested TF QKD system, Alice/Bob employes a laser source with the linewidth of less than $0.1~\kilo\hertz$, whose output optical power ranges from $12~\milli\watt$ to $40~\milli\watt$. Alice and Bob maintain a fixed light-frequency difference of 112 MHz between their local light and the reference light using an OPLL, whose locking bandwidth is $112 \pm 30~\mega\hertz$ and the locking time is about $0.2~\micro\second$. Once the OPLL is locked, the response time of the FLC to control AOM is between $3~\micro\second$-$10~\micro\second$.
This identical frequency difference ensures that Alice and Bob have consistent frequencies with each other. Without any attack, as shown in~\cref*{fig:detectpd1}(a), a single peak of the heterodyne frequency appears at $112~\mega\hertz$ with a maximum power of $-9.14~\deci\bel\milli$. Under the wavelength-switching attack conducted by Eve as depicted in the red area in~\cref*{fig:tfqkdscheme}(a), where there is a rapid change in the wavelength of the reference light, the heterodyne frequency is no longer fixed at $112~\mega\hertz$. We consider a set of attack parameters where there is maximum wavelength change ($f_{\Delta}=30~\mega\hertz$) and switching speed within the response range of FLC to control AOM ($r_s=100~\kilo\hertz$). As shown in~\cref*{fig:detectpd1}(b), under this attack, two peaks appear at frequencies $112~\mega\hertz$ and $142~\mega\hertz$, respectively. The difference between these two peaks depends on the introduced attack frequency ($f_{\Delta}$). Additionally, it can be observed from~\cref*{fig:detectpd1}(b) that these two frequency peaks have power values of $-15.14~\deci\bel\milli$ and $-16.75~\deci\bel\milli$, respectively. The measured peak power of heterodyne frequency under this attack still allows for proper functioning of OPLL since FLC only requires a minimum power greater than or equal to $-30~\deci\bel\milli$, which remains satisfied even under this specific attack.

In a TF-class QKD system, Charlie provides the reference light to Alice and Bob and performs interferometric measurements on the quantum signals they send back, as depicted by the blue area in~\cref*{fig:tfqkdscheme}(a). Ensuring stable interference is crucial for successful interferometric measurement in order to achieve high visibility of quantum state detection. This step plays a critical role in establishing secure key generation between Alice and Bob within the TF-class QKD system. Therefore, it is essential to verify the stability of interference under wavelength-switching attacks. To maintain stable interference despite such attacks, Eve manipulates the wavelength of the reference laser while simultaneously affecting Alice's and Bob's lasers. The key lies in selecting an appropriate switching rate ($r_s$) that achieves stable interference. \Cref*{fig:tfqkdscheme}(b) illustrates Eve's selection of four switching rates ($r_s$) as attacking parameters. The black and red solid lines represent optical power measured by PD3 and PD4 at two output arms of Charlie's interference setup respectively. When there is no attack ($r_s=0$), Charlie measures smooth and stable interference over time. However, slight fluctuations are observed when Alice's and Bob's wavelengths are switched at rates of either $100$~$\kilo\hertz$ or $200$~$\kilo\hertz$, indicating potential disturbance caused by these attacks.

The slight fluctuations begin to decrease at the switching rate of $400$~$\kilo\hertz$. In this scenario, the response time of OPLL is surpassed by $r_s$, rendering the OPLL incapable of promptly transmitting the frequency command. Consequently, two potential outcomes emerge: if the FLC fails to transmit the frequency shift command to Alice's and Bob's AOM, system security remains uncompromised; however, successful transmission of the frequency shift command by FLC to Alice's and Bob's AOM triggers an effective wavelength-switching attack. Subsequently, in following iterations of feedback loop, a sustained frequency shift can be maintained resulting in more stable interference compared to cases where $r_s=100~\kilo\hertz$ and $r_s=200~\kilo\hertz$, as illustrated in~\cref{fig:tfqkdscheme}(b). Nevertheless, it should be noted that continuous operation of the frequency shift command on the AOM is not guaranteed. Gradually, this command is transferred from the AOM to PZT by FLC which reduces both wavelength shifting applied by AOM and optical intensity simultaneously. In summary, attacks at a switching rate of $400$~$\kilo\hertz$ do not always exhibit stability or controllability.

The conducted experiment illustrates that, notwithstanding slight variations, the overall influence of a wavelength-switching attack on Charlie's measured interference in a TF-class QKD system remains constrained. As a result, Eve can strategically adapt her wavelength switching rate according to real-time conditions for her benefit while minimally disturbing the functionality of the TF QKD system.

\subsubsection{Gain of average photon number under attack}

Consequently, we conduct a comprehensive investigation into the augmentation of the mean-photon number in Alice's and Bob's quantum states when subjected to a wavelength-switching attack. The magnitude of this augmentation serves as an indicator for quantifying the information intercepted by Eve from Alice and Bob, thereby providing crucial insights into potential eavesdropping on secure keys.

Experimental data in~\cref*{fig:tfqkdscheme}(c) illustrates the output power of Alice (Bob) with and without a wavelength-switching attack. In the absence of an attack, the optical power output of Alice or Bob remains relatively stable, as depicted in the leftmost chart in~\cref*{fig:tfqkdscheme}(c). At this juncture, the driving voltage of the AOM is $8.9~\volt$, indicated by the black dot in~\cref{fig:attenuation}(b). As $f_{\Delta}$ increases, both Alice's and Bob's optical intensity experiences fluctuations during different phases of the switching cycle shown by the rest of charts in~\cref*{fig:tfqkdscheme}(c). This behavior arises from energy storage characteristics inherent to $\rm T_eO_2$ material used in AOMs~\cite{oliveira2023}. With increasing $f_{\Delta}$ values, ultrasonic power also rises correspondingly, leading to an increase in AOM voltage (indicated by red upward-pointing triangle marker in~\cref{fig:attenuation}(b)). Conversely, when ultrasonic power decreases, so does the driving voltage of AOM (as shown by red downward-pointing triangle marker in ~\cref{fig:attenuation}(b)). It is noteworthy that increased optical power resulting from attacks represents a gain exceeding designed mean photon number for both Alice's and Bob's quantum states. The gain value can be calculated as increased optical power divided by designed optical power. Higher gain values imply greater benefits for Eve. From ~\cref*{fig:tfqkdscheme}(c), it can be inferred that under varying $f_{\Delta}$ values such as $1~\mega\hertz$, $3~\mega\hertz$, $9~\mega\hertz$, $20~\mega\hertz$, and $30~\mega\hertz$ respectively; output gains for either Alice or Bob are found to be 0.86\%, 1.63\%, 3.19\%, 4.04\% and 4.35\%.

\begin{figure}[htbp]
	\centering
	\includegraphics[width=0.48\textwidth]{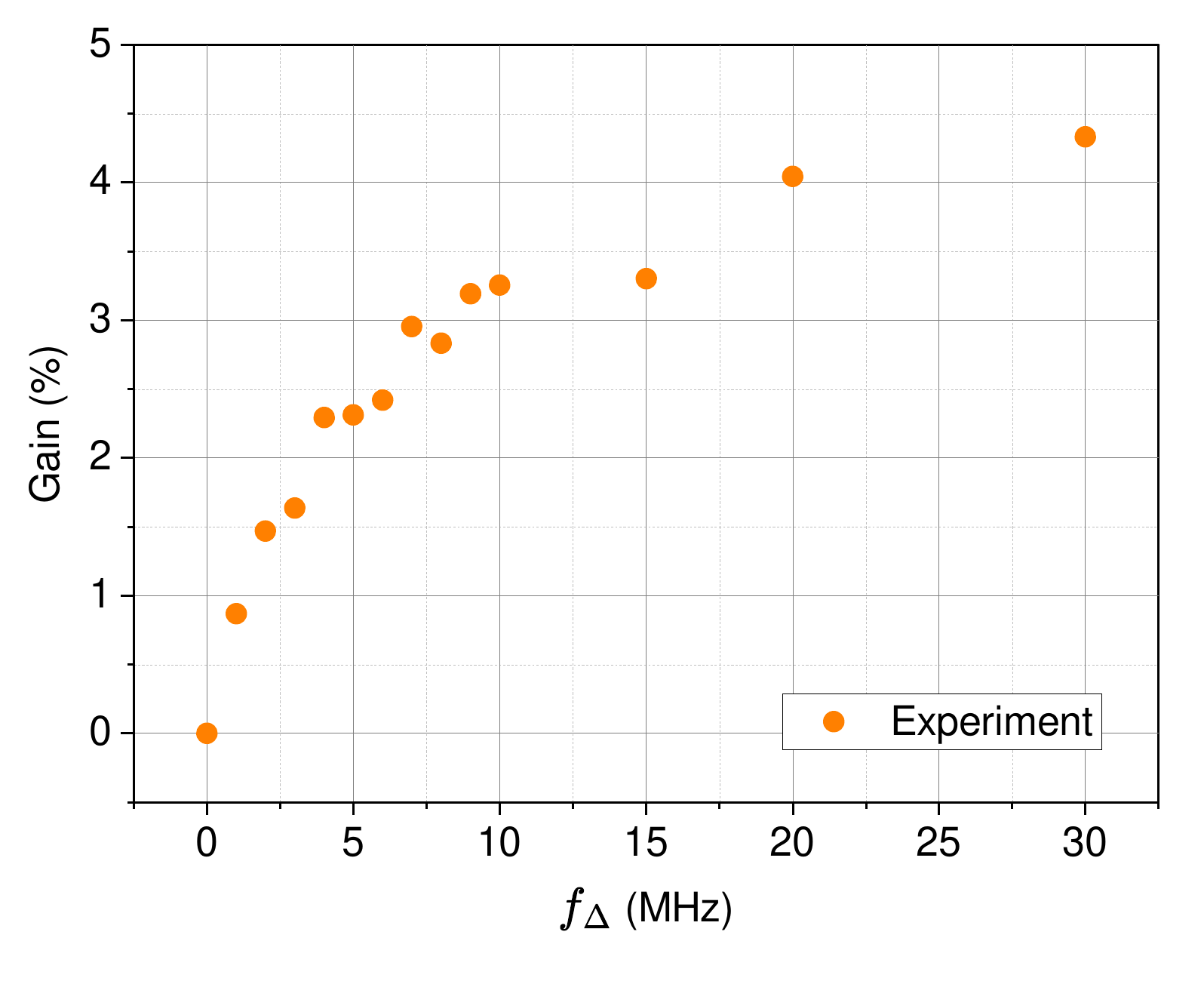}
	\caption{
		The experimental results demonstrated the average photon number gain of the OPLL for each source apparatus in the TF-class QKD system under the wavelength-switching attack.
	}
	\label{fig:gain}
\end{figure}

The correlation between the shifted wavelength under attack and the gain in mean photon number of emitted quantum states in the TF-class QKD system is illustrated in~\cref{fig:gain}. The experimentally measured values after conversion are denoted by yellow markers. The gain in mean photon number of the TF-class QKD system exhibits a rapid increase when $f_\Delta$ is less than or equal to 10~MHz. After that, the rate of increase slows down. The gain in mean photon number is maximized at $f_\Delta = 30~\mega\hertz$, beyond which the OPLL is no longer working in the locking status. Therefore, we limit our frequency shift to a maximum of $30~\mega\hertz$ and apply it simultaneously to the OPLLs of both Alice and Bob during the experiment. As a result, the mean photon number of Alice/Bob is increased by 4.35\%, and the overall intensity gain is 8.7\% for the whole TF QKD system as defined in Ref.~\cite{lucamarini2018}.

\subsection{Effect on the security of TF-class QKD}
\label{sec:discuss}

The experiment demonstrates that the wavelength-switching attack can enhance the output power of Alice's and Bob's laser source, resulting in a factor of $g = G+1 > 1$ for the mean photon numbers of all quantum states. Assuming that Alice's and Bob's photon numbers still follow a Poisson distribution, all $\mu$ values become ${\mu}' = g\mu$ under the wavelength-switching attack. Consequently, it is necessary to theoretically evaluate the impact of this attack on the original TF QKD system. In this study, we present a comprehensive security analysis of the TF QKD system and extend our analysis to include the sending-or-not-sending TF QKD (SNS-TF-QKD) system.

\subsubsection{Security analysis on TF QKD}

In the TF QKD system, Alice and Bob generate phase-randomized optical fields, which are transmitted to Charlie through a quantum channel. Subsequently, Charlie performs the interferometric measurement and announces an event only if a single-photon detector is clicked. The event will be considered valid for generating a secure key if Alice and Bob select the same random phase. To evaluate the potential vulnerability of wavelength-switching attacks, we consider typical implementations where Alice/Bob uses three intensity settings $\mu_{a,b} \in \left \{ \mu_0/2,\mu_1/2,\mu_2/2 \right \} $, with $\mu_0=0.4$, $\mu_1=10^{-2}$, and $\mu_2=10^{-4}$. Here, $\mu_0$ represents the signal state while $\mu_1$ and $\mu_2$ represent decoy states. In scenarios with a realistic number of transmitted signals approaching asymptotic limits, it is possible to establish a lower bound on the secret key rate~\cite{lucamarini2018}.

\begin{equation}
\begin{aligned}
R_{L} = \frac{d}{M} \left \{ Q_{1,L}^{\mu_0}[1-H_{2}(e_{1,U})]-fQ^{\mu_0}H_{2}(E^{\mu_0}) \right \},
\label{eq:tfqkd}
\end{aligned}
\end{equation}
where $d=0.5$ is the duty cycle between the classical and the quantum modalities, $M=16$ is the value of phase slices, $Q_{1,L}^{\mu_0}$ denotes a lower bound on the single-photon yield, $e_{1,U}$ denotes an upper bound on the phase error rate, $f=1.16$ is the efficiency of error correction, $Q_U$ represents the overall experimentally observed gain, $E_u$ is the overall experimentally observed quantum bit error rate (QBER), and $H_2(x) = -x\log_2(x)-(1-x)\log_2(1-x)$ is the binary Shannon entropy function.

\begin{figure}[htbp]
\centering
\includegraphics[width=0.5\textwidth]{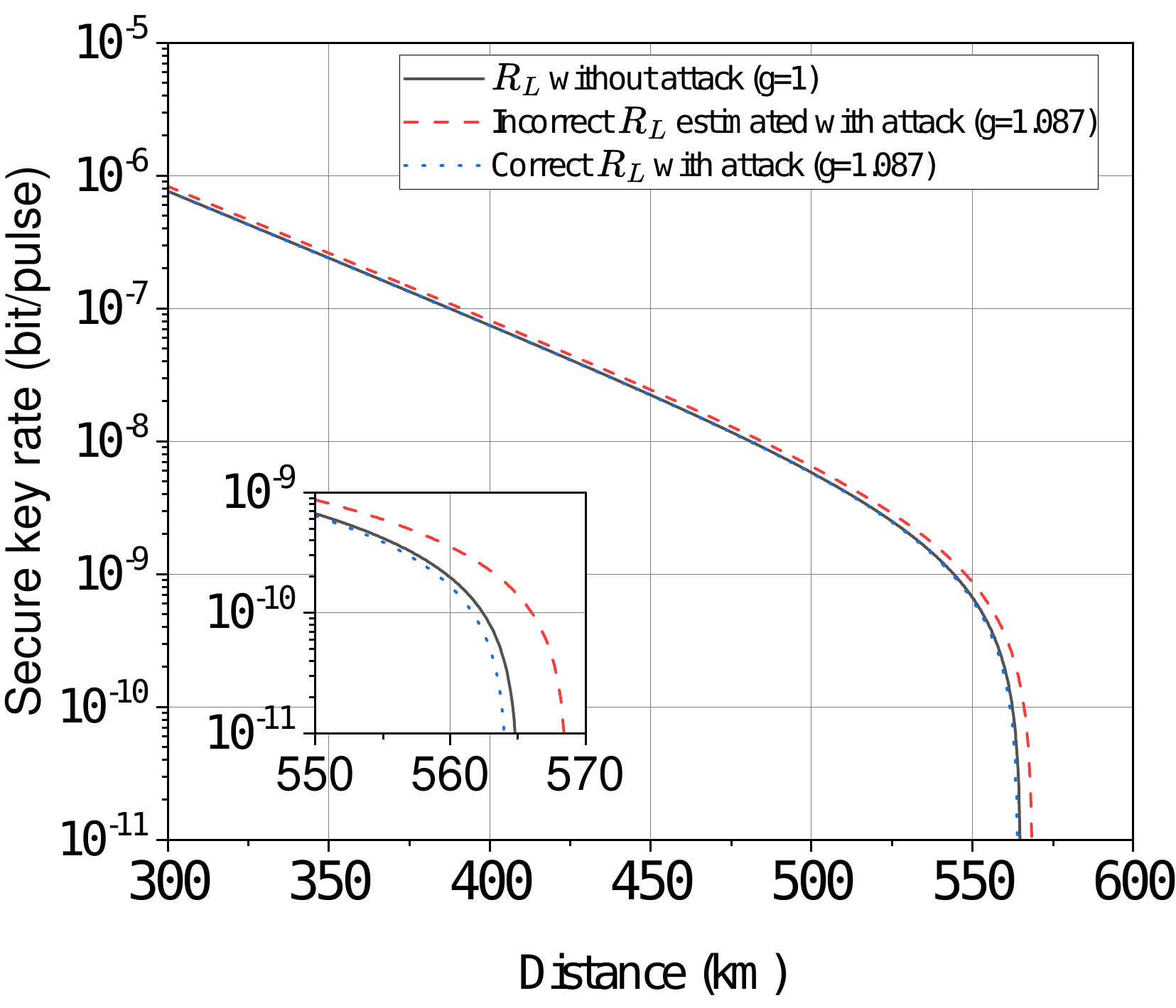}
\caption{
	The secure key rate is plotted as a function of distance for TF QKD, considering two different values of the multiplicative factor: $g=1$ and $g=1.087$. 
}
\label{fig:tfqkd}
\end{figure} 

\begin{figure}[htbp]
\centering
\includegraphics[width=0.5\textwidth]{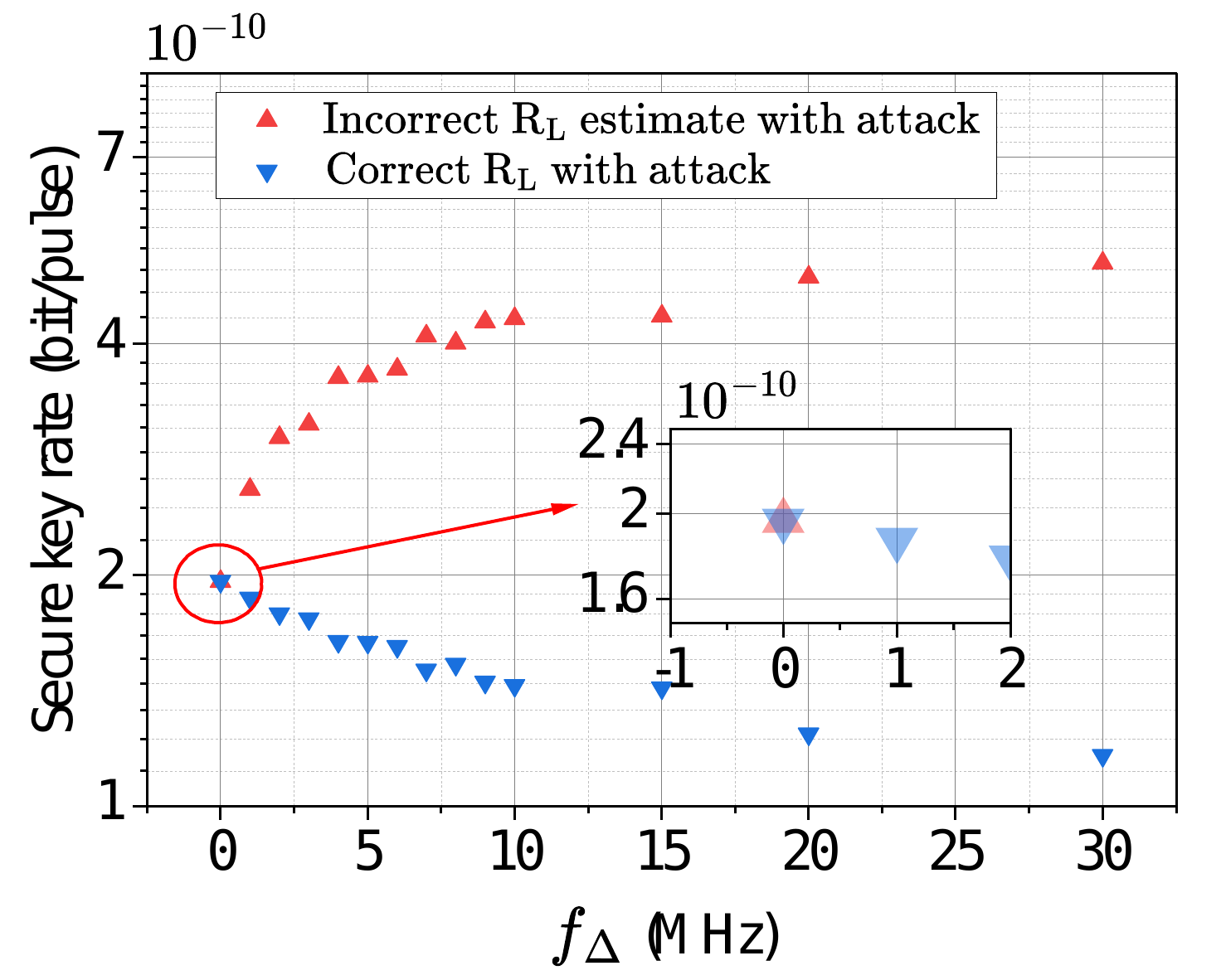}
\caption{
	The secure key rate is plotted as a function of the parameter $f_\Delta$ for a fixed distance of $560~\kilo\metre$ in the TF QKD system.
}
\label{fig:tfqkdfre}
\end{figure} 

To estimate $Q_{1,L}$, $Q^{\mu}$, $e_{1,U}$ and $E_\mu$, one can use analytical or numerical tools. Here we have that

\begin{equation}
\begin{aligned}
  Q_{1,L}^{\mu_0} &\ge \mu_0 e^{-\mu_0} Y_{1,L},\\
  Q^{\mu_{0,1,2}} &= 1-(1-p_d)^2e^{-\eta\mu_{0,1,2}},
\end{aligned}
\label{eq:q1}
\end{equation}
with $\eta = \eta_{det} \times 10^{-\frac{\alpha L/2}{10}}$, $\eta_{det}=30\%$ is total detection efficiency, $\alpha = 0.2~\rm{dB/km}$ is the linear attenuation of a standard optical fibre, and $L$ is the distance.

\begin{equation}
\begin{aligned}
  e_{1,U} \le \frac{E^{\mu_1}Q^{\mu_1}e^{\mu_1}-E^{\mu_2}Q^{\mu_2}e^{\mu_2}}{(\mu_1-\mu_2)Y_{1,L}},
\end{aligned}
\label{eq:e1}
\end{equation}
where 

\begin{equation}
\begin{aligned}
  Y_{1,L} \ge & \frac{{\mu_0}^2}{\mu_0({\mu_0}{\mu_1}-{\mu_0}{\mu_2}-{\mu_1}^2+{\mu_2}^2)}\\
  & \times \left [ Q^{\mu_1}e^{\mu_1}-Q^{\mu_2}e^{\mu_2}-\frac{{\mu_1}^2-{\mu_2}^2}{{\mu_0}^{2}}(Q^{\mu_0}e^{\mu_0}-Y_{0,L}) \right ], \\
  Y_{0,L} \ge & \frac{{\mu_1Q^{\mu_2}e^{\mu_2}} - {\mu_2Q^{\mu_1}e^{\mu_1}}}{{\mu_1}-{\mu_2}}.
\end{aligned}
\label{eq:y1}
\end{equation}

For the error rates,
\begin{equation}
\begin{aligned}
  E^{\mu_{0,1,2}} &= \frac{1}{2}+\frac{1-p_d}{2Q^{\mu_{0,1,2}}}\left[e^{-\mu_{0,1,2}\eta(1-e_{opt}-E_m)} -e^{-\mu_{0,1,2}\eta(e_{opt}+E_m)} \right],\\
  E_{m} &= \frac{1}{2} - \frac{\sin(2\pi/M)}{4\pi/M},
\end{aligned}
\label{eq:eu}
\end{equation}
with $e_{opt} = 3\%$ is the channel optical error rate. In the presence of the wavelength-switching attack, Alice and Bob estimate $e_{1,U}$, $Y_{0,L}$, and $Y_{1,L}$ using \cref{eq:e1} and \cref{eq:y1}, but now calculate $E^{\mu_{0,1,2}}$ and $Q^{\mu_{0,1,2}}$ using the experimentally observed parameters $\mu'_{0,1,2} = g\times\mu_{0,1,2}$.
We conduct a simulation analysis using parameters corresponding to a typical experiment~\cite{wang2022} to illustrate the security threat posed by a wavelength-switching attack on the TF QKD system. The results with and without an attack on the TF QKD system are depicted in~\cref{fig:tfqkd}. The black solid line represents the lower bound $R_L$ as defined by \cref{eq:tfqkd} under no attack conditions. Subsequently, we simulate the degradation of security bounds due to Eve's wavelength-switching attack. Specifically, in~\cref{fig:tfqkd}, the red-dashed and blue-dotted lines respectively indicate values of incorrectly estimated ($R_L$) by Alice and Bob when $g=1.087$ in the presence of an attack. As shown in~\cref{fig:tfqkd} compared to no-attack scenario represented by black solid line, the secure key rate ($R_L$) given by red-dashed line is significantly higher. In other words, Alice and Bob overestimate their secure key rate in response to an attack. The gap between these two curves indicates that Eve can eavesdrop on a portion of secure key rate without being detected while eventually deducing shared secure key between Alice and Bob.

In~\cref{fig:tfqkdfre}, we present the impact of the wavelength-switching factor $f_\Delta$ on the lower bound of secure key rate at a fixed distance ($560~\kilo\metre$) and $g=1.087$. The graph illustrates both incorrect and correct estimates of secure key rates for the TF QKD system under attack, represented by upward triangles and downward triangles respectively. In this scenario, an increase in parameter $f_\Delta$ results in a higher mean photon number for quantum states prepared by Alice and Bob. This consistently elevated mean photon number leads to an overestimation of the lower bound $R_L$ of key rate, as indicated by the red-square markers, thereby widening the gap between overestimated and correct key rates illustrated by blue-dot markers. To summarize, wavelength-switching attacks cause Alice and Bob to overestimate their key rate, compromising the security of a TF QKD system.

\subsubsection{Security analysis on SNS-TF-QKD}
Next, we consider the case of SNS-TF-QKD~\cite{wang2018}. Similar to the previous example, it is assumed that both Alice and Bob employ three distinct intensities, namely $\mu_0$, $\mu_1$, and $\mu_2$. In the finite-key case, the lower bound for the secure key rate can be derived from Ref.~\cite{jiang2019,hu2022,liu2023}.

\begin{equation}
\begin{aligned}
  R_L= \frac{1}{N} \left \{ n_{1}\left[ 1-H_2(e_1^{ph})\right] - fn_{t}H_2(E_z) \right \}-\gamma,
\label{eq:snstfqkd}
\end{aligned}
\end{equation}
where $N$  represents the finite size, $n_1$ denotes the lower bound of the number of survived untagged bits, $e_1^{ph}$ gives the upper bound of the phase-flip error rate for those survived untagged bits, $f$ is the error correction inefficiency, $n_t$ is the number of survived untagged bits, $E_z$ is the corresponding bit-flip error rate in those survived bits, and $\gamma$, which takes logarithmically small values, is for security considerations with finite-data size and advanced decoy state analysis.

To evaluate \cref{eq:snstfqkd}, Alice and Bob need to calculate the lower bound of the expected value of the counting rate of untagged photons

\begin{equation}
\begin{aligned}
  S_{1,L} \ge & \frac{1}{2\mu_1\mu_2\left ( \mu_2 - \mu_1 \right ) } \\
           & \times [ {\mu_2}^2 e^{\mu_1} (S_{01} + S_{10}) \\
           & - {\mu_1}^2 e^{\mu_2} (S_{02} + S_{20}) -2({\mu_2}^2-{\mu_1}^2)S_{00} ],
\end{aligned}
\end{equation}
where $S_{jk} \left ( jk=\left \{ 00,10,01,20,02 \right \} \right ) $ denotes the counting rate of source $jk$ \cite{jiang2019}. 
In the wavelength-switching attack, the parameters $S_{jk}$ should be calculated based on the experimentally obtainable data $\mu'_0$, $\mu'_1$, and $\mu'_2$.
According to Ref.~\cite{jiang2019,liu2023}, $n_1$ and $n_t$ can be obtained from $S_{1,L}$, $E_z$ can be calculated based on $S_{jk}$, and $e_1^{ph}$ can be estimated from $e_{1,U}$. 
The upper bound of the expected value of the phase-flip error rate satisfies \cite{yu2019}

\begin{equation}
\begin{aligned}
  e_{1,U} &\le \frac{T_\Delta - 0.5e^{-2\mu_1}S_{00}}{2\mu_1e^{-2\mu_1}S_{1,L}},
\end{aligned}
\end{equation}
where $T_\Delta$ is the effective rate of detectors. 
Furthermore, we can obtain a much lower QBER $E_z$ in those survived bits by applying active odd-parity pairing~\cite{xu2020a,jiang2020,jiang2021} to the above method. 

\begin{figure}[htbp]
\centering
\includegraphics[width=0.5\textwidth]{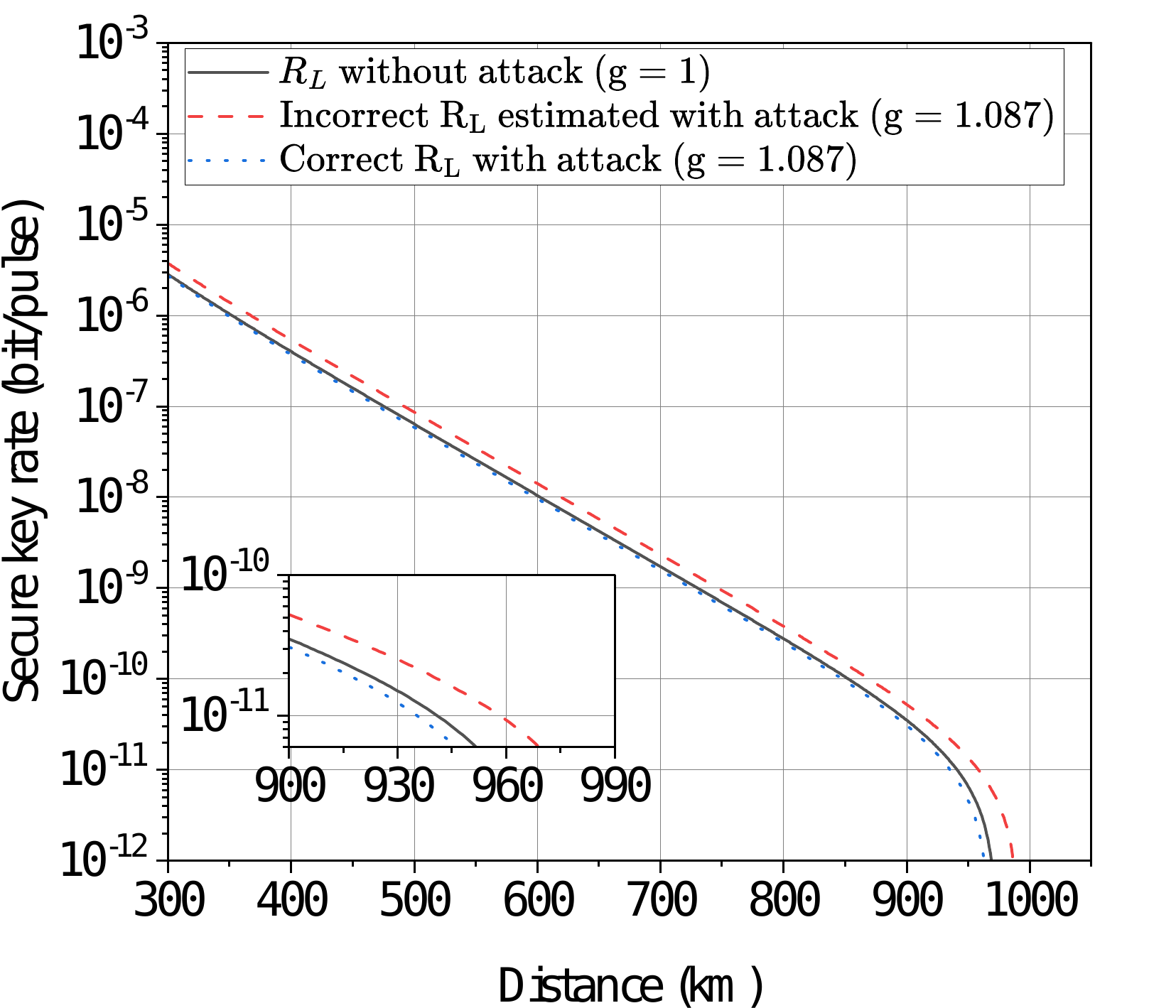}
\caption{
	The secure key rate as a function of distance for SNS-TF-QKD is observed to be $1$ and $1.087$ for two different values of the multiplicative factor $g$. The simulation parameters utilized in the calculations are as follows. $N = 5.4 \times 10^{14}$, $P_d = 1.4 \times 10^{-11}$, $f = 1.16$, $\alpha = 0.157$, and $\eta_{det} = 0.6$.
}
\label{fig:snstfqkd}
\end{figure} 

We adopt a methodology similar to that employed in the preceding subsection to assess $R_L$ under the presence of a wavelength-switching attack. We employ parameters from a representative case~\cite{liu2023} for simulation purposes. For simplicity, we assume Eve executes a symmetric attack by injecting identical wavelength-switching light into both Alice and Bob. The resulting lower bound on the key rate is depicted in~\cref{fig:snstfqkd}. In this example, we consider two potential values for the gain factor, namely $g = 1$ and $g = 1.087$. The scenario with $g = 1$ corresponds to an absence of attacks. Evidently, the incorrect secure key estimated by Alice and Bob in the presence of an attack surpasses the correct value.

\section{Discussion}
\label{sec:discussion}

\subsection{Countermeasures}
\label{sec:countermeasures}

Our experimental results indicate that the TF QKD employed by the external reference-light-based OPLL is vulnerable to wavelength-switching attacks. An eavesdropper can manipulate Alice's and Bob's output photon numbers by altering the wavelength of the reference light, compromising the practical security of TF-class QKD systems. Consequently, it is imperative to develop countermeasures against such attack.

The observed variation in the intensity of quantum states is attributed to the change in optical insertion loss. This change is mainly because the AOM receives an uncharacterized frequency modulation and insufficient driving voltage when the FLC adjusts it to match the reference light's wavelength switching as analyzed in~\cref{subsec:aom}. Being immune to the wavelength-switching attack, we propose a calibration procedure for the AOM before system operation to minimize the insertion loss due to driving voltage. In step 1, the driving voltage of the AOM is scanned under a fixed modulation frequency to determine the optimal voltage value, $V_o$, corresponding to the minimum insertion loss at this certain modulation frequency. In step 2, the modulation frequency of the AOM is scanned with $V_o$ found in Step 1, obtaining the optimal frequency $F_o$, corresponding to the further minimized insertion loss. Afterwards, $V_o$ and $F_o$ are set as the basic parameters for the FLC to control the AOM. This process ensures that the wavelength-switching attack will only increase the insertion loss of the OPLL, thus mitigating the associated security risks.

An additional countermeasure could be involving monitoring devices to detect eavesdropping. For example, during heterodyne measurements, photodetectors PD1 and PD2 might be used to identify multiple peaks in the spectrum, potentially indicating a wavelength-switching attack.
Moreover, by monitoring the intensity with PD5 and PD6 in the source stations, the system can monitor the intensity fluctuation of Alice's and Bob's output light. These fluctuations can be indicative of the potential wavelength-switching attack.
From another aspect of view, inspired by the fully passive QKD protocol \cite{hu2023,lu2023}, the vulnerability related to wavelength-switching attacks might be mitigated if the wavelength/optical field of a laser source can be passively modulated or locked.

An alternative approach for Alice and Bob to counter the wavelength-switching attack is to modify the configuration of the TF QKD system by employing a local frequency reference. In this way, Alice/Bob in the TF-class QKD system does not open a door for any untrusted party, fundamentally eliminating the security threat introduced by the external reference light. Recent research has proposed a practical method that utilizes the saturated absorption spectroscopy of acetylene as an absolute reference, thereby eliminating the need for external reference light in TF-class QKD implementation \cite{chen2024}. This method indeed can prevent Eve from using external reference light to attack the TF-class QKD system. However, integrating acetylene as a new component may introduce potential risks that necessitate further investigation.

\subsection{Conclusion}
\label{sec:conclusion}

The TF-class QKD has been demonstrated to surpass the fundamental rate-distance limit, which necessitates the generation of identical optical fields by two independent lasers located at separate positions for Alice and Bob. In current implementations, a third-party reference light is commonly employed to lock the wavelengths of the lasers in order to achieve matching. Although this reference-light-locking structure does not directly contribute to key generation, it introduces a potential vulnerability that can be exploited by an eavesdropper, posing a security risk to the devices used by Alice and Bob. This study unveils a wavelength-switching attack on the TF-class QKD system employing an optical OPLL structure. Experimental results demonstrate that Eve can exploit this attack to reduce attenuation in the AOM of the TF-class QKD system, causing optical power from Alice and Bob to exceed intended levels. Importantly, this attack challenges fundamental assumptions in TF-class QKD protocols regarding mean photon number of prepared quantum states and well-protected source stations. Despite altering wavelengths of lights emitted by Alice and Bob, these hacked source units remain consistent, enabling stable interference at Charlie's location. Theoretical analysis conducted on both TF QKD protocol and SNS-TF-QKD protocol indicates that the wavelength-switching attack could significantly compromise security in implementation of TF-class QKD - legitimate users overestimate secure key rates when facing such attacks. Since this attack is due to imperfect calibration in the external-reference-locking configuration, we propose countermeasures via loss-minimization calibration or using no-external-reference-light scheme~\cite{zhou2023,zhou2023a,Li2023,zhu2023,chen2024}. 
The possibility of this attack on various OPLLs schems shall be experimentally verified in the future.
Additionally, the practical security of other variants of TF-QKD, such as phase-matching QKD~\cite{ma2018}, shall be investigated as future study.

\section{Methods}
\label{sec:methods}

\subsection{Model of AOM}
\label{App:AOM}

\begin{figure}[htbp]
\centering
\includegraphics[width=0.5\textwidth]{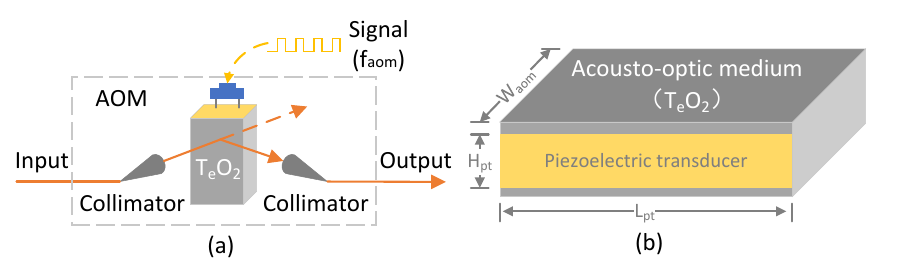}
\caption{
  (a) Schematic diagram of the AOM.
  (b) Acousto-optic medium and piezoelectric transducer. 
}
\label{fig:aom_medium}
\end{figure}

The AOM device operates by converting electrical energy into ultrasonic vibrations. Adjusting the modulation frequency in this process directly affects the electrical energy, thereby influencing the wavelength and intensity of light within an acoustic-optic medium. To provide a clearer explanation, we have employed a simplified model as depicted in~\cref{fig:aom_medium}(a) to establish a direct correlation between modulation frequency and AOM attenuation values in this specific context. This approach facilitates a more accessible comprehension of the interrelationship among key factors in our study.

\begin{equation}
\begin{aligned}
  I_{aom} = -10\log_{10}{\frac{P_{aom}^{\rm{out}}}{P_{aom}^{\rm{in}}}} + \delta_{aom}.
\end{aligned}
\end{equation}

The input optical intensity of the AOM, denoted as $P_{aom}^{\rm{in}}$, is set at $6.22 \milli\watt$ in the given scenario. The output optical intensity, $P_{aom}^{\rm{out}}$, can be expressed as $\eta_{aom} P_{aom}^{\rm{in}}$, where $\eta_{aom}$ represents the diffraction efficiency of the acousto-optic medium used in the AOM. Moreover, to simplify the simulation, we introduce $\delta_{aom}$ as a fixed value of $3.57~\deci\bel$ to account for other attenuations. It should be noted that this parameter may vary experimentally due to factors such as coupling ratio and practical considerations. Mathematically, the diffraction efficiency ($\eta_{aom}$) can be mathematically expressed as follows \cite{yu2011}:

\begin{equation}
\begin{aligned}
  \eta_{aom} = \sin^{2}\left ( \frac{\pi}{\lambda} \sqrt{\frac{M_{aom,2}L_{\rm{pt}}P_{aom,a}}{2H_{\rm{pt}}}} \right ),
\end{aligned}
\end{equation}
where $\lambda = 1550~\nano\meter$ is the wavelength of light. $M_{aom,2}$ is the acousto-optic figure of merit for the crystal.

$H_{\rm{pt}}$ and $L_{\rm{pt}}$ correspond to the thickness and length of the piezoelectric transducer, respectively, as shown in~\cref{fig:aom_medium}(b). $P_{aom,a}$ is the ultrasonic power, which can be written as

\begin{equation}
\begin{aligned}
  P_{aom,a} =& \frac{F_{aom}^2}{Z_{aom,M}},\\
  F_{aom} =& \frac{f_{aom}}{f_{aom,0}},
\end{aligned}
\label{eq:paom}
\end{equation}
where $F_{aom}$ is the relative frequency. $f_{aom}$ is the modulation frequency. $f_{aom,0}$ is the half wavelength frequency of the piezoelectric layer. $Z_{aom,M}$ is the relative acoustic impedance of the acousto-optic interaction medium.

The AOM depicted in~\cref{fig:aom_medium}(b) is composed of glass-ceramics and features a $\rm T_eO_2$ composition, which is well-known for its ability to minimize scattering losses at the interfaces between glass and crystal. This characteristic makes it an exceptional host material for precipitating crystalline phases \cite{oliveira2023}. The value of $M_{aom,2}$ for this medium has been measured at $34.7 \times 10^{-15} \rm{s^3/kg}$~\cite{yu2011}. Moreover, this medium exhibits energy storage properties that make it highly suitable for advanced electronic applications such as electric microchips, hybrid vehicles, and pulse power systems~\cite{tian2019,du2020,khalf2021}. In essence, when subjected to increasing ultrasonic power $P_{aom,a}$, $\rm T_eO_2$ rapidly charges up while effectively suppressing excessive energy; conversely, it releases stored energy during power decrease.

\backmatter

\bmhead{Acknowledgements}
We thank X.-B. W., Q. Z., and C. J. for valuable discussions.
This work was funded by the Innovation Program for Quantum Science and Technology (2021ZD0300704) and National Natural Science Foundation of China (No. 62371459 and No. 62061136011).

\bmhead{Disclosures}
The authors declare no conflicts of interest.

\bmhead{Data availability}
The data that support the findings of this study are available from the corresponding
author upon request.

\bmhead{Code availability}
The codes used for numerical analysis are available from the corresponding author upon request.

\bmhead{Author contributions}
Q.P., J.P., and A.H. conducted the experiment. T.X., D.W., and Y.W. support the experimental conduct. Q.P., Y.L., and A.H. analyzed the data. Q.P. and A.H. wrote the paper with input from all authors. A.H. and Y.L. supervised the project.


\bibliographystyle{naturemag}
\bibliography{library}

\end{document}